\documentclass[11pt,a4paper]{article}
\usepackage{jheppub}
\relax

\def\bes{\begin{eqnarray}}
 \def\ees{\end{eqnarray}}
\def\be{\begin{equation}}
\def\ee{\end{equation}}
\def\bs{\begin{subequations}}
\def\es{\end{subequations}}
\newcommand{\een}{\end{subequations}}
\newcommand{\ben}{\begin{subequations}}
\newcommand{\beq}{\begin{eqalignno}}
\newcommand{\eeq}{\end{eqalignno}}

 \def\kx{\kappa}
 \def\Lx{\Lambda}

 \def\kb{{\bar{\kappa}}}

\def\rb{{\bf r}}
\def\mut{{\tilde{\mu}}}
\def\nut{{\tilde{\nu}}}
\def\kxt{{\tilde{\kx}}}
\def\kbt{{\tilde{\kb}}}

\def \lta {\mathrel{\vcenter
     {\hbox{$<$}\nointerlineskip\hbox{$\sim$}}}}
\def \gta {\mathrel{\vcenter
     {\hbox{$>$}\nointerlineskip\hbox{$\sim$}}}}

\title{
The renormalization of fluctuating branes,\\ the Galileon and asymptotic safety
}
\author[1]{A. Codello,}
\author[2]{N. Tetradis}
\author[3]{and O. Zanusso}

\affiliation[1]{SISSA, 
Via Bonomea 265, 34136 Trieste, Italy}
\affiliation[2]{Nuclear and Particle Physics Sector, Department of Physics, \\
University of Athens, Zographou 15784, Greece}
\affiliation[3]{PRISMA Cluster of Excellence and Institute of Physics (THEP),\\
University of Mainz, Staudingerweg 7, D-55099 Mainz, Germany}

\emailAdd{codello@sissa.it}
\emailAdd{ntetrad@phys.uoa.gr}
\emailAdd{zanusso@thep.physik.uni-mainz.de}

\abstract{
We consider the renormalization of $d$-dimensional hypersurfaces (branes) embedded in flat $(d+1)$-dimensional space. 
We parametrize the truncated effective action in terms of geometric invariants built from the
extrinsic and intrinsic curvatures. 
We study the renormalization-group running of the couplings and explore the
fixed-point structure. We find evidence for an ultraviolet fixed point similar to the one underlying the
asymptotic-safety scenario of gravity.
We also examine whether the structure of the Galileon theory, which can be reproduced in the nonrelativistic limit, 
is preserved at the quantum level.
}

\preprint{MZ-TH/12-54}

\begin{document}

\maketitle

\section{Introduction}\label{intro}

Scalar field theories are ubiquitous in physics, describing a plethora of classical and quantum systems.
Because of their relative simplicity, they have often been used as a testing ground for new ideas or techniques.
The action of a scalar field is usually assumed to contain a standard kinetic term, 
especially when quantum or statistical fluctuations of the system are studied. 
The inclusion of higher-derivative terms may lead to two pathologies: a) The presence of derivatives higher than the
second in the equation of motion results in the appearance of modes with negative norm, characterized as ghosts. 
b) The higher-derivative terms are perturbatively nonrenormalizable and the theory loses predictivity. 

We are interested in the systematic study of quantum corrections in scalar field theories with higher-derivative terms. As we have 
mentioned, such theories are
in general nonrenormalizable in the perturbative sense. However, their scale dependence can be
studied through the exact renormalization group (ERG). Their renormalizability may result from the presence of 
a nonperturbative fixed point. The main drawback of the ERG approach
is that the integration of the flow equation for the scale-dependent effective action can be achieved only for
truncated versions of the action. However, it is still possible to check the reliability of the predictions by expanding the 
truncation scheme and examining their stability. This procedure has been applied to scalar theories with a general potential and a
standard kinetic term, leading
to an accurate determination of nontrivial quantities, such as critical exponents \cite{exp}. 
The precision can be improved by going to higher orders of the derivative expansion \cite{canet}.

The theories we consider in this work describe hypersurfaces, which we term branes, embedded 
in a higher-dimensional flat spacetime, to which we refer as bulk spacetime. The leading contribution to the action is 
given by the volume swept by the brane, expressed in terms of the induced metric.
It is invariant under arbitrary changes of the brane worldvolume coordinates. We can fix this gauge freedom by
identifying the brane coordinates with certain bulk coordinates. 
This choice is usually characterized as 
the static gauge. The remaining bulk coordinates become scalar fields of the worldvolume theory.
More complicated terms can also be included in the effective action. 
The crucial property that makes this class of theories interesting is that the effective action can be expressed in 
terms of geometric quantities, such as the intrinsic and extrinsic curvatures of the hypersurface. In the static gauge
these can be written in terms of the scalar fields and their derivatives. 
In this way we obtain a higher-derivative scalar theory with a particular structure.
The ERG flow of the scalar theory describing 
two-dimensional membranes has been considered in ref.~\cite{codello}.
Here we present the generalization to a $d$-dimensional brane, embedded in a $(d+1)$-dimensional bulk. 

Scalar field theories with derivative interactions have been considered extensively during the last years in the context of 
particle physics and cosmology under a variety of names, such as: $k$-essence \cite{kessence}, Dirac-Born-Infeld (DBI) 
inflation \cite{dbin},
the Dvali-Gabadadze-Porrati (DGP) model \cite{dgp} in the decoupling limit and the Galileon \cite{galileon}, 
scalar-tensor models with kinetic gravity braiding \cite{braiding}, etc. All these theories are constructed so that
the equation of motion does not contain field derivatives higher than the second,
even though a large or infinite series of derivative terms can be present in the 
action. In this way, ghost fields do not appear in the spectrum.  
The most general scalar-tensor theory with this property was constructed a long time ago \cite{horndeski}, 
and rediscovered recently. It is characterized as the generalized Galileon (see ref.~\cite{genegal} and references therein).

The absence of derivatives higher than the second in the equation of motion is not protected by some underlying symmetry.
For example, for the Galileon theory it is known that quantum corrections generate terms that destroy this property.
The one-loop corrections computed through dimensional regularization 
induce a term $\phi \Box^4 \phi$ in the effective action \cite{quantum}. 
It  is still possible to consider the Galileon as a consistent quantum theory at low energies, for which such a term is 
subleading. The main motivation for our study stems from the wish to understand the issue of quantum corrections for
such derivative theories through the ERG approach. 

The connection between the Galileon and the brane picture that we discussed above is provided by 
ref.~\cite{dbigal}, which shows that the Galileon theory can be reproduced 
in the nonrelativistic limit by considering 
the effective action for the position modulus of a probe brane within a five-dimensional bulk. Derivatives 
higher than the second can be avoided by employing only Lovelock invariants in the geometric picture.
In this work we consider a truncation of the brane effective action that takes into account the
lowest-order geometric invariants. Some of these reproduce the structure of the Galileon theory \cite{galileon}. 
On the other hand, our truncation
includes a contribution involving the extrinsic curvature of the brane 
that does not have an analogue in the Galileon theory, as it would induce a field derivative higher than the second 
in the equation of motion. We examine how this term scales under quantum corrections and whether it is consistent
to assume that it does not appear in the effective action. 

Our study has another very interesting spinoff. If the contributions from the extrinsic curvature are omitted,
the ERG flow can be expressed as the evolution of an effective Newton's constant and the cosmological constant. 
The picture is similar to that obtained in ERG studies of gravity, in which the metric is considered as the fundamental field.
The $\beta$-functions display a fixed-point structure that is analogous to that associated with asymptotic safety \cite{asysaf}. 
Thus we obtain a very useful testing ground for a concept that could provide the UV completion of gravity.

In the following section we establish our notation and summarize the correspondence between the brane and Galileon theories.
In section \ref{sec3} we introduce the effective action that we consider and the flow equation that describes its evolution.
In section \ref{sec4} we derive the $\beta$-functions for the couplings of the theory and discuss the effect of 
quantum corrections on the structure of the brane and Galileon theories. In section \ref{sec5} we analyze
the fixed-point structure for a consistent truncation that preserves only the cosmological-constant and Einstein
terms. We discuss the analogy with the asymptotic-safety scenario of gravity. In the final section we present a summary
and our conclusions. 
The ERG formalism has been developed for field theories in Euclidean space. For this reason we assume the 
analytic continuation to imaginary time throughout the paper. 

\section{Brane dynamics and the Galileon}\label{sec2}

Following ref.~\cite{dbigal},
we summarize briefly the connection between the dynamics of fluctuating branes and the Galileon theory.
The connection has been established for a four-dimensional brane embedded in five-dimensional flat space. 
The induced metric in the static gauge is $g_{\mu\nu}=\eta_{\mu\nu}+\partial_\mu \pi \, \partial_\nu \pi$, where
$\pi$ denotes the extra coordinate of the bulk space. We preserve the notation $\eta_{\mu\nu}$
even though we use imaginary time and the bulk metric is Euclidean. 
The induced extrinsic curvature is 
$K_{\mu\nu}=-\partial_\mu\partial_\nu\pi/\sqrt{1+(\partial\pi)^2}$. We denote its trace by $K$.
The leading terms in the brane effective action
are
\begin{eqnarray}
S_\mu&=&\mu\int d^4x  \sqrt{g}=\mu\int d^4 x \sqrt{1+(\partial \pi)^2}
\label{sl} \\
S_\nu&=&\nu\int d^4x  \sqrt{g}\, K=-\nu\int d^4x\, \left([\Pi]-\gamma^2[\phi]\right)
\label{sn} \\
S_\kb&=&\frac{\kb}{2}\int d^4x  \sqrt{g}\, R=\frac{\kb}{2}\int d^4x \, \gamma
\left([\Pi]^2-[\Pi^2] +2\gamma^2([\phi^2]-[\Pi][\phi]) \right),
\label{skb} 
\end{eqnarray}
where $\gamma=1/\sqrt{g}=1/\sqrt{1+(\partial \pi)^2}$.
We have adopted the notation of ref.~\cite{dbigal}, with $\Pi_{\mu\nu}=\partial_\mu\partial_\nu \pi$ and square brackets
representing the trace (with respect to $\eta_{\mu\nu}$) of a tensor. Also, we denote 
$[\phi^n]\equiv\partial \pi\cdot \Pi^n \cdot \partial\pi$, 
so that $[\phi]= \partial^\mu\pi \, \partial_\mu\partial_\nu\pi\, \partial^\nu\pi$. 
The field $\pi$ has mass dimension $-1$, as it corresponds to a spatial coordinate. It can be given a more conventional
mass dimension through multiplication with the appropriate power of the fundamental energy scale $M$ of the theory. 
We implicitly assume that all other scales are expressed in terms of $M$, which is effectively set equal to 1. 
The couplings $\mu$, $\nu$, $\kb$ correspond to the effective four-dimensional cosmological constant, the five-dimensional
Planck scale $M_5^3$ and the four-dimensional Planck scale $M_4^2$, respectively.

The effective action of the Galileon theory can be obtained in the nonrelativistic limit $(\partial \pi)^2 \ll 1$. It must be
noted, however, that terms with second derivatives of the field, such as $\pi\Box \pi$, are not assumed to be small (here $\Box=\eta^{\mu\nu}\partial_\mu\partial_\nu$). 
If total derivatives are neglected, the integrants of the leading terms in the expansion of (\ref{sl})--(\ref{skb}) are
proportional to $(\partial \pi)^2$. In this way, one obtains three of the terms appearing in the 
action of the Galileon theory \cite{dbigal}. The term of highest order in this theory can be obtained by including in the
brane action the Gibbons-Hawking-York term associated with the Gauss-Bonnet term of five-dimensional gravity. 
We omit this term in the truncated effective action that we consider, as it complicates significantly the
study of the renormalization of the theory. Its effect  will be the focus of future work. In the context of
asymptotic safety boundary terms have been considered in \cite{Becker:2012js}.

The first Gauss-Codazzi equation gives $R=K^2-K^{\mu\nu}K_{\mu\nu}$. This relation indicates that the 
truncation of the effective brane action that includes a term $\sim R$ should also include a
term $\sim K^2$. On the other hand, such a term must be excluded if the equation of motion is assumed not 
to contain field derivatives higher than the second. Its absence 
cannot be enforced by some underlying symmetry, and quantum corrections may introduce it even if it is omitted
in the tree-level action. In order to study its role in the renormalized theory we include in our
truncated action the contribution
\be
S_\kx=\frac{\kx}{2}\int d^4x  \sqrt{g}\, K^2=\frac{\kx}{2} \int d^4x\,\gamma \left([\Pi]-\gamma^2[\phi]\right)^2.
\label{sk} \ee 
In the limit $(\partial \pi)^2 \ll 1$, the contribution $[\Pi]^2$ included in
this term generates in the integrant a leading contribution $\sim \pi \Box^2 \pi$. 
A term $K^{\mu\nu}K_{\mu\nu}$ in the Lagrangian density would produce a contribution $[\Pi^2]$, which
would again become $\sim \pi \Box^2 \pi$ in the nonrelativistic limit.
The two leading contributions cancel in $R=K^2-K^{\mu\nu}K_{\mu\nu}$, so that the structure of the 
Galileon is generated. On the other hand, if quantum corrections spoil the cancellation, the Galileon theory is not reproduced. 

It is worth pointing out that the term $S_{\nu}$ can be omitted if we assume the discrete symmetry $\pi \to -\pi$. 
The same symmetry would eliminate the higher-order contribution related to the Gauss-Bonnet term of the 
bulk theory. For a probe brane the presence in the action of terms odd in the extrinsic curvature indicates an asymmetry 
in the fluctuations on either side of the brane. The origin of such terms is not obvious, unless the bulk space is not homogeneous or the
brane is viewed as its boundary.  
These considerations indicate that it seems more natural to include the contributions (\ref{sl}), (\ref{skb}), (\ref{sk}) 
in a consistent quantum theory than the ones that reproduce the Galileon theory. The
terms (\ref{sl}), (\ref{skb}), (\ref{sk}) form 
the basis for the study of the renormalization of two-dimensional fluid membranes (see ref.~\cite{codello} and references therein).

\section{Flow equation}\label{sec3}

The focus of our study is the evolution of the scale-dependent effective action 
\be
\Gamma_k=\int d^d x  \sqrt{g}\left(
\mu_k+\nu_k K+\frac{\kx_k}{2}K^2+\frac{\kb_k}{2}R
\right),
\label{avac} \ee
with the various invariants expressed through the field $\pi$.
We have included the contributions (\ref{sl})-(\ref{sk}) discussed in the previous section, but we now
assume that the various couplings depend on the running energy scale $k$. The action describes the dynamics of
a $d$-dimensional brane embedded in ($d+1$)-dimensional flat space. We use imaginary time, so that the bulk metric
is Euclidean.

The formal treatment of the action (\ref{avac}) can be carried out through the ERG.
We first introduce the scale $k$ by adding to the action a term $R_k(q^2)$ in momentum space, so that fluctuations of
the field with characteristic momenta $q^2\lta k^2$ are cut off \cite{wetterich,exp}.
We subsequently introduce sources and define the generating functional for the connected
Green functions. Through a Legendre transformation we
obtain the generating functional for the 1PI Green functions, from which we subtract the regulating term involving $R_k(q^2)$. 
In this way we obtain the scale-dependent effective action
$\Gamma_k[\pi]$.
The procedure results in the effective integration of the fluctuations with 
$q^2\gta k^2$. The theory is assumed to possess a fundamental high-energy cutoff $M$, so that 
$\Gamma_k$ is identified with the 
bare action $S$ for $k=M$. For $k=0$ we obtain the standard effective action. 
The means for calculating $\Gamma_k$ from $S$ is provided by the exact flow equation \cite{wetterich}
\be
\partial_t \Gamma_k[\pi]=\frac{1}{2}{\rm Tr} \frac{\partial_t R_k(-\Box)}{\Gamma_k^{(2)}[\pi]+R_k(-\Box)},
\label{flow} \ee
where we have reverted to position space and defined $t=\ln k$. Here $\Gamma_k^{(2)}[\pi]$ denotes 
the second functional derivative of the action with respect to the field.
The rhs of the above equation receives contributions only from fluctuations with characteristic momenta $q^2\simeq k^2$. In this sense, the high-energy cutoff $M$ is only a formal element in the definition of $\Gamma_k$. It can be replaced by a UV fixed point in the flow of $\Gamma_k$. 

When gauge symmetries, such as the reparametrization invariance of the brane worldvolume theory, are present the 
definition of $\Gamma_k$ is more involved. We shall not present the details here, and we refer the reader to 
refs.~\cite{asysaf} for the case of gravity, and to ref.~\cite{codello} for the case of brane reparametrization invariance.
In the scale-dependent action the reparametrization invariance is implemented through the use of the background field method.
The brane position is determined by the embedding function $\rb=(x^\mu,\pi)$ and the induced metric is given by $g_{\mu\nu}=\partial_{\mu}\rb\cdot\partial_{\nu}\rb=\eta_{\mu\nu}+\partial_\mu \pi \, \partial_\nu \pi$.  
We parametrize the fluctuations around a background configuration $\rb$ as $\rb+\delta \rb$. In the
static gauge that we have adopted, we have $\delta \rb=\delta\pi {\bf n}$, where ${\bf n}$ is the unit vector normal to the
brane and $\delta\pi$ is the fluctuating field. 
The cutoff $R_k(\Delta)$ is constructed by means of the operator $\Delta=-g^{\mu\nu}\nabla_{\mu} \nabla_{\nu}$,
where the induced metric $g_{\mu\nu}$ and covariant derivatives $\nabla_{\mu}$ compatible with it are expressed
in terms of the (background) field $\pi$.
The scale-dependent (background) effective action $\Gamma_k[\left\langle\delta\pi \right\rangle;\pi]$ 
depends both on the (background) field $\pi$ and the expectation value
$\left\langle \delta\pi \right\rangle$ of the fluctuating field. It is constructed so as to be invariant under
reparametrizations of the background, while
the expectation value $\left\langle\delta\pi \right\rangle$ has to transform covariantly on a given background $\pi$.

The exact flow equation has the form
\be
\partial_t \Gamma_k[\left\langle \delta\pi \right\rangle;\pi]=\frac{1}{2}{\rm Tr} \frac{\partial_t R_k(\Delta)}{\Gamma_k^{(2;0)}[\left\langle \delta\pi \right\rangle;\pi]
+R_k(\Delta)}.
\label{flowg} \ee
In the limit $\left\langle\delta\pi \right\rangle\to 0$
the background invariance is promoted to full reparametrization invariance.
The effective action can be identified with $\Gamma_k[\pi]\equiv\Gamma_k[0;\pi]$.
The difficulty we have to face is that the flow equation (\ref{flowg}) is not a closed relation for $\Gamma_k[0;\pi]$.
It becomes closed if we make the ansatz $\Gamma_k^{(2;0)}[0;\pi]=\Gamma_k^{(0;2)}[0;\pi]\equiv\Gamma_k^{(2)}[\pi] $.
We obtain eq.~(\ref{flow}), where now the d'Alembertian $-\Box$ is replaced by 
the d'Alembertian $\Delta$ constructed with the full induced metric.
As truncation ansatz for the effective action $\Gamma_k[\pi]$ we choose eq. (\ref{avac}), which is reparametrization 
invariant by construction. In this way, the invariance is preserved by the evolution, even though the 
full dependence of the functional  $\Gamma_k[\left\langle\delta\pi \right\rangle;\pi]$ on the two
fields $\pi$ and $\left\langle\delta\pi \right\rangle$ is not taken into account. 
(For example, this functional in principle includes 
a separate wavefunction renormalization for the fluctuation field $\left\langle\delta\pi \right\rangle$.) 

There is a more intuitive, albeit less rigorous, way to generate the flow equation. 
The one-loop correction to a tree-level action of the form (\ref{avac}) is proportional to the logarithm of the fluctuation
determinant around a given background. In order to compute it, we employ the static gauge and expand the field
as $\rb+\delta\pi {\bf n}$, keeping only the terms quadratic in $\delta\pi$. The resulting expression depends on
the various couplings appearing in (\ref{avac}). These are now the bare ones and have no $k$-dependence. 
The contribution of fluctuations with characteristic momenta below a given scale $k$ can be excluded 
if we add to the Lagrangian density a term $\sim \delta\pi R_k(\Delta) \delta\pi$. 
It must be kept in mind that the theory (\ref{avac}) has a geometric origin, which must be preserved even when
we employ the static gauge and express the action in terms of the field $\pi$. The cutoff must be 
constructed in a way consistent with this property. This can be achieved if we construct the d'Alembertian employing the
full induced metric, expressed in terms of $\pi$. A ``renormalization-group improvement" of the effective action can be
achieved by taking its logarithmic derivative with respect to $k$ and substituting the running couplings, which
are $k$-dependent, for the bare
ones. The resulting expression is the flow equation we discussed above. 

\section{$\beta$-functions}\label{sec4}

Extracting information from the flow equation requires an appropriate parametrization and truncation of 
the scale-dependent effective action. For this purpose we employ the truncation (\ref{avac}). In order to calculate 
the trace in the
rhs of the flow equation we need the second functional derivative of (\ref{avac}) on an arbitrary background. 
We find 
\be
\Gamma_k^{(2)}[\pi]= \kx_k\Delta^2 +\mu_k\Delta +V^{\mu\nu}\nabla_\mu \nabla_\nu+U+{\cal O}(K^4,\nabla K),
\label{hes} \ee
where 
\begin{eqnarray}
V^{\mu\nu}&=&2\nu_k\left( K^{\mu\nu}-K g^{\mu\nu}\right) +\kx_k \left[
-\frac{1}{2}\left(3 K^2-4 K^{\rho\sigma}K_{\rho\sigma}  \right)g^{\mu\nu}+2K K^{\mu\nu}
\right] 
\nonumber \\
&&+\kb_k\left[
\left( R^{\mu\nu}-\frac{1}{2}R\, g^{\mu\nu}\right)-\left( K^2-K^{\rho\sigma}K_{\rho\sigma} \right)g^{\mu\nu}+2K K^{\mu\nu}
+2K^{\mu\sigma}K^\nu_{~\sigma} \right]
\label{vmunu}\\
U&=&\mu_k\left(K^2-K^{\rho \sigma} K_{\rho\sigma} \right)
\label{umunu}
\end{eqnarray}
and the covariant derivatives are evaluated with the full induced metric.
The first Gauss-Codazzi equation allows us to express $K^2-K^{\rho\sigma}K_{\rho\sigma}$ in terms of $R$ in the above
expressions. A similar simplification can be carried for $K^{\mu\sigma}K^\nu_{~\sigma}$. 
However, we have preserved the expression in a form similar to that given in ref.~\cite{codello} for the two-dimensional brane.

We substitute the above expressions in the rhs of the flow equation and expand the denominator in powers of the curvatures.
The trace of the resulting terms can be computed through the heat kernel expansion, as described in \cite{rgmachine}. The details of this procedure have
been presented in ref.~\cite{codello} for the case $d=2$ and we do not repeat them here. 
We insert the truncation (\ref{avac}) in the lhs of the flow equation and match the contributions that involve the 
same curvature invariants on both sides of the equation. In this way we obtain the $\beta$-functions for the various 
couplings. They are 
\begin{eqnarray}
\partial_t\mu_k&=&\frac{1}{(4\pi)^{d/2}} \frac{1}{2}\, Q_{\frac{d}{2}}\left[G_k\, \partial_t R_k  \right]
\label{bmuk} \\
\partial_t\nu_k&=&-\frac{1}{(4\pi)^{d/2}} \frac{d-1}{2}\, Q_{\frac{d}{2}+1}\left[G^2_k\, \partial_t R_k  \right] \nu_k
\label{bnuk} \\
\partial_t\kx_k&=&\frac{1}{(4\pi)^{d/2}} \Biggl\{
\frac{d+4}{4}\, Q_{\frac{d}{2}+1}\left[G^2_k\, \partial_t R_k  \right] \kx_k
+(d^2-1)\, Q_{\frac{d}{2}+2}\left[G^3_k\, \partial_t R_k  \right] \nu^2_k
\Biggr\}
\label{bkxk} \\
\partial_t\kb_k&=&\frac{1}{(4\pi)^{d/2}} \Biggr\{
\frac{1}{6}\, Q_{\frac{d}{2}-1}\left[G_k\, \partial_t R_k  \right] 
-Q_{\frac{d}{2}}\left[G^2_k\, \partial_t R_k  \right] \mu_k
 -2\,Q_{\frac{d}{2}+2}\left[G^3_k\, \partial_t R_k  \right] \nu^2_k \Biggr.
\nonumber \\
&&\Biggl.
-d\, Q_{\frac{d}{2}+1}\left[G^2_k\, \partial_t R_k  \right] \kx_k
-\frac{3(d-2)}{4}\, Q_{\frac{d}{2}+1}\left[G^2_k\, \partial_t R_k  \right] \kb_k
\Biggr\}.
\label{bkbk} \end{eqnarray} 
The regularized propagators are 
\be
G_k(z)=\frac{1}{\kx_k z^2+\mu_k z + R_k(z)},
\label{prop} \ee
while the $Q$-functionals are defined as
\begin{eqnarray}
Q_n[f]&=&\frac{1}{\Gamma(n)}\int_0^\infty dz\, z^{n-1} f(z)~~~~~~~~~~~~n>0
\nonumber \\
Q_n[f]&=&(-1)^n f^{(n)}(0)~~~~~~~~~~~~~~~~~~~~~~~~n\leq 0.
\end{eqnarray}

Some qualitative properties of the evolution are immediately apparent. The $\beta$-function of $\nu_k$ vanishes
for $\nu_k=0$. This is an expected result, as setting $\nu_k=0$ in the tree-level action induces the discrete
symmetry $\pi\to -\pi$, which protects this value at the quantum level as well. 
 For $\nu_k \not= 0$, which is a necessary assumption in order to reproduce the Galileon theory in the nonrelativistic limit,
the $\beta$-function of $\kx_k$ does not vanish. It is apparent from eq.~(\ref{bkxk}) that a contribution 
$\sim \nu_k^2 \, K^2$ is induced through quantum fluctuations. In the nonrelativistic limit 
a term $\sim \nu_k^2 \, \pi \Box^2 \pi$ will be generated, which is not present in the Galileon theory.
A similar phenomenon occurs for a scalar field coupled to gravity \cite{Eichhorn:2012va}.
On the other hand, the analysis of the one-loop corrections to the Galileon through the use
of dimensional regularization shows that the lowest-order
induced term is $\sim \nu_k^2 \,\pi \Box^4 \pi$ \cite{quantum}.

In order to understand this point we need to make contact with perturbation theory. 
With the appropriate approximations,
the $\beta$-functions (\ref{bmuk})-(\ref{bkbk}) can reproduce standard perturbative results. For 
$\nu_k=\kx_k=0$ the scale-dependent effective action (\ref{avac}) has the same structure as
Einstein gravity with a cosmological constant \cite{asysaf}. 
The one-loop contribution to the cosmological constant can be obtained if we set $\mu_k=1$ in the rhs of 
eq.~(\ref{bmuk}). This is the bare value of this parameter that leads to a canonically normalized kinetic
term when $\sqrt{g}$ is expanded in powers of $(\partial \pi)^2$ and the leading term is retained. 
Independently of the choice of cutoff function $R_k(q^2)$, we obtain (with $z=q^2$)
\be
\partial_t \mu_k=\partial_t \left[\frac{1}{2}\int \frac{d^dq}{(2\pi)^d}\ln(q^2+R_k(q^2)) \right].
\label{oneloopmu} \ee
The trivial integration of this equation for $k$ in the range $[0,M]$ 
reproduces the one-loop contribution
to the vacuum energy arising from the quantum fluctuations of a single massless mode in a theory with a fundamental
high-energy cutoff $\sim M$. It must be noted that in the brane theory the renormalization of the
cosmological constant is the same as that of the leading kinetic term at low energies. This is obvious from
the form of the propagator (\ref{prop}), in which $z=q^2$ is multiplied by $\mu_k$. As a result the
field $\pi$ has a large anomalous dimension. 

We can obtain the one-loop correction to $\kx_k$ in a similar fashion, by substituting the bare couplings for the running ones
in the rhs of eq.~(\ref{bkxk}). We assume that the bare theory does not contain a term $\sim K^2$ and the 
kinetic term is canonically normalized. With these assumptions we can set $\mu_k=1$ and $\kx_k=0$ in the 
rhs of  (\ref{bkxk}) and replace $\nu_k$ by a constant value $\nu_M$. We obtain
\be
\partial_t \kx_k=-\frac{2(d^2-1)}{d(d+2)} \nu_M^2\,
\partial_t \left[ \int \frac{d^dq}{(2\pi)^d}\frac{q^4}{(q^2+R_k(q^2))^2} \right].
\label{oneloopkx} \ee
The integration of this equation for $k$ in the range $[0,M]$ results in a momentum integral with a quartic divergence for $d=4$, 
which is
cut off by a high-energy scale $\sim M$. This quantum correction would not be visible if dimensional regularization was used. 
On the other hand, the regularization with an explicit cutoff, such as the one employed in the context of the ERG, 
picks up corrections with 
possible quadratic or quartic divergences. The correction of eq.~(\ref{oneloopkx}) induces a term
$\sim \nu_M^2 \, K^2$, which in the nonrelativistic limit becomes 
$\sim \nu_M^2 \, \pi \Box^2 \pi$.
This term is not present in the Galileon theory, and is 
of a lower order than the term $\sim \nu_M^2 \, \pi \Box^4 \pi$
expected from an analysis based on dimensional regularization. 
We emphasize that this conclusion does not require a specific 
choice of the cutoff function $R_k(z)$, and thus is ERG-scheme independent. 

A cross-check of the $\beta$-functions (\ref{bmuk})-(\ref{bkbk}) can be obtained if we set $\mu_k=\nu_k=0$. For $d=2$ the resulting 
theory can describe two-dimensional fluid membranes in three-dimensional space. The couplings $\kx_k$ and $\kb_k$ correspond
to the bending and Gaussian rigidities. The $\beta$-functions of these couplings were computed in 
ref.~\cite{codello}. 
They agree with those derived through perturbation theory 
\cite{twod} if the anomalous dimension of the fluctuating field is set to zero.

\section{Fixed points and asymptotic safety}\label{sec5}

Explicit expressions for the $\beta$-functions can be obtained for specific forms of the cutoff function
$R_k(z)$. 
The results are particularly simple for the choice 
\be
R_k(z)=\left[\kx_k (k^4-z^2)+\mu_k (k^2-z)\right]\theta(k^2-z).
\label{cutoff} \ee
Despite its unconventional form, the cutoff function generates the required behavior for the effective propagator $1/G_k(q^2)$:
For $z=q^2>k^2$ the effective propagator is the perturbative one ($1/G_k(z)=\kx_k z^2+\mu_k z$), so that
the corresponding fluctuations remain unaffected by the presence of the cutoff. For
$z<k^2$, $1/G_k(z)$ is finite and constant ($1/G_k(z)=\kx_k k^4+\mu_k k^2$) and the low-energy fluctuations are 
suppressed. It has been verified through several studies that, when these criteria are fulfilled, 
the predictions obtained in the limit $k\to 0$ are independent of the specific form of $R_k(z)$ \cite{exp}. 

For $R_k(z)$ given by eq.~(\ref{cutoff}) the function $f(z)$ in eqs.~(\ref{bmuk})-(\ref{bkbk}) has
the general form $f(z)=[G_k(z)]^m \partial_t R_k(z)$, with $m$ a positive integer. Our cutoff choice leads to
the appearance of terms $\sim \partial_t\kx_k, \partial_t \mu_k$ in  $\partial_t R_k(z)$. For $\kx_k=0$, the term
$\partial_t \mu_k$ would correspond to the anomalous dimension of the field. 
We shall neglect these contributions in our analysis, as they
are not expected to affect the qualitative features of the evolution. They must be included, however, if
quantitative precision is required.

For $n\geq 0$, the $Q$-functionals become
\be
Q_n[f]=\frac{k^{2n}}{\Gamma(n+1)}f(0),
\label{Qn} \ee
with
\be
f(0)=\left(\frac{1}{\kx_k k^4+\mu_k k^2} \right)^m\left(4\kx_k k^4+2\mu_k k^2 \right).
\label{f0} \ee
The evolution equations (\ref{bmuk})-(\ref{bkbk}) take the form
\begin{eqnarray}
\partial_t\mu_k&=&\frac{k^d}{(4\pi)^{d/2}\Gamma\left(\frac{d}{2}+1\right)} \frac{2\kx_k k^2+\mu_k}{\kx_k k^2+\mu_k}
\label{bbmuk} \\
\partial_t\nu_k&=&- \frac{k^d}{(4\pi)^{d/2}\Gamma\left(\frac{d}{2}+2\right)}(d-1)
\frac{(2\kx_k k^2+\mu_k)\nu_k}{(\kx_k k^2+\mu_k)^2}
\label{bbnuk} \\
\partial_t\kx_k&=& \frac{2k^d}{(4\pi)^{d/2}\Gamma\left(\frac{d}{2}+2\right)}
\Biggl\{ \frac{d+4}{4}\frac{(2\kx_k k^2+\mu_k)\kx_k}{(\kx_k k^2+\mu_k)^2}
+ \frac{4(d^2-1)}{d+4} \frac{(2\kx_k k^2+\mu_k)\nu^2_k}{(\kx_k k^2+\mu_k)^3}  \Biggr\}
\label{bbkxk} \\
\partial_t\kb_k&=& \frac{k^d}{(4\pi)^{d/2}\Gamma\left(\frac{d}{2}+2\right)} \Biggr\{
\frac{d(d+2)}{12}\frac{2\kx_k k^2+\mu_k}{(\kx_k k^2+\mu_k)k^2}
-\frac{8}{d+4} \frac{(2\kx_k k^2+\mu_k)\nu^2_k}{(\kx_k k^2+\mu_k)^3}
\Biggr.
\nonumber \\
&&\Biggl.
-\left[ (d+2)\frac{\mu_k}{k^2}+2d\kx_k+\frac{3(d-2)}{2}\kb_k\right]
 \frac{2\kx_k k^2+\mu_k}{(\kx_k k^2+\mu_k)^2}
\Biggr\}.
\label{bbkbk} \end{eqnarray} 
The structure of the above equations is typical of the ERG, with terms involving various powers of the
effective propagator appearing in the $\beta$-functions. The class of theories that we are considering
involves only generalized kinetic terms. For this reason couplings such as $\kx_k$, $\mu_k$ that multiply  
the leading terms appear often in the denominator in the $\beta$-functions. 

As the theory has a geometric origin, the fundamental field
$\pi$ has mass dimension $-1$ because it corresponds to a spatial coordinate. As we have already mentioned, 
it can be given a more conventional
mass dimension through multiplication with the appropriate power of the fundamental energy scale $M$ of the theory. 
Throughout the paper we assume that all scales are expressed in terms of $M$. The scaling dimensions of the
various couplings can be deduced from eqs.~(\ref{bbmuk})-(\ref{bbkbk}) if we remove the explicit factors of
$k$ through the appropriate redefinitions. If we define 
\be
\mu_k=k^{d}\mut_k,~~~~
 \nu_k=k^{d-1}\nut_k,~~~~
 \kx_k=k^{d-2}\kxt_k,~~~~
 \kb_k=k^{d-2}\kbt_k,
\label{dimless} \ee
eqs.~(\ref{bbmuk})-(\ref{bbkbk}) become
\begin{eqnarray}
\partial_t\mut_k&=&-d\mut_k+\frac{1}{(4\pi)^{d/2}\Gamma\left(\frac{d}{2}+1\right)}\frac{2\kxt_k +\mut_k}{\kxt_k +\mut_k}
\label{bbmukt} \\
\partial_t\nut_k&=&-(d-1)\nut_k- \frac{1}{(4\pi)^{d/2}\Gamma\left(\frac{d}{2}+2\right)}(d-1)
\frac{(2\kxt_k+\mut_k)\nut_k}{(\kxt_k+\mut_k)^2}
\label{bbnukt} \\
\partial_t\kxt_k&=& -(d-2)\kxt_k+\frac{2}{(4\pi)^{d/2}\Gamma\left(\frac{d}{2}+2\right)}
\Biggl\{ \frac{d+4}{4}\frac{(2\kxt_k +\mut_k)\kxt_k}{(\kxt_k +\mut_k)^2}
+ \frac{4(d^2-1)}{d+4} \frac{(2\kxt_k +\mut_k)\nut^2_k}{(\kxt_k +\mut_k)^3}  \Biggr\}
\nonumber \\
&~&
\label{bbkxkt} \\
\partial_t\kbt_k&=& -(d-2)\kbt_k +\frac{1}{(4\pi)^{d/2}\Gamma\left(\frac{d}{2}+2\right)} \Biggr\{
\frac{d(d+2)}{12}\frac{2\kxt_k +\mut_k}{\kxt_k+\mut_k}
-\frac{8}{d+4} \frac{(2\kxt_k +\mut_k)\nut^2_k}{(\kxt_k +\mut_k)^3}
\Biggr.
\nonumber \\
&&\Biggl.
-\left[ (d+2)\mut_k+2d\kxt_k+\frac{3(d-2)}{2}\kbt_k\right]
 \frac{2\kxt_k +\mut_k}{(\kxt_k +\mut_k)^2}
\Biggr\}.
\label{bbkbkt} \end{eqnarray} 
This is the most convenient form of the evolution equations for the determination of their fixed points. 

As a first check we can compute the $\beta$-functions of $\kx_k$, $\kb_k$ for two-dimensional fluid membranes.
In the membrane theory the volume (now area) term is considered subleading. This means that we can get the relevant
equations by
setting $d=2$, $\mu_k=\nu_k=0$ in eqs.~(\ref{bbkxk}), (\ref{bbkbk}). We obtain
\be
\partial_t \kx_k=\frac{3}{4\pi},~~~~~~~~~~~\partial_t\kb_k=-\frac{5}{6\pi}.
\label{fluid} \ee
These expressions reproduce the results of refs.~\cite{codello,twod} for the renormalization of the
bending and Gaussian rigidities of fluctuating membranes in a three-dimensional bulk space. It must be pointed out, however,
that the relation $\mu_k=0$ is not consistent with eq.~(\ref{bbmuk}), which becomes 
\be
\partial_t \mu_k=\frac{k^2}{2\pi}
\label{fluidmu} \ee 
for $d=2$, $\mu_k=0$. Neglecting the area term can be viewed only as a low-energy approximation.
Setting $\mu_k=\nu_k=0$ in eqs.~(\ref{bbkxk}), (\ref{bbkbk}) provides a generalization of the evolution for branes of 
arbitrary dimensionality.

An important point, which we have already discussed in the previous section, is the stability of the conditions $\nu_k=0$ and
$\kx_k=0$ under quantum corrections. The first one is expected to be stable, as it is protected by the 
symmetry $\pi\to-\pi$. The evolution equation (\ref{bbnuk}) explicitly demonstrates that $\partial_t\nu_k$ vanishes 
for $\nu_k=0$. On the other hand the condition $\kx_k=0$ does not enhance the symmetry of the action and
is not expected to survive at the quantum level. Eq. (\ref{bbkxk}) indicates that corrections $\sim\nu_k^2$ are generated
for $\kx_k$ under renormalization. It is noteworthy that, if we set $\nu_k=0$, we have $\partial_t\kx_k=0$ for $\kx_k=0$ and
$\mu_k\not=0$. We believe that this is an accidental feature. Notice also that the $\beta$-function does not vanish if we first 
set $\mu_k=0$ and then take the limit $\kx_k\to 0$. 

The analysis of the fixed points of the system of equations (\ref{bbmukt})-(\ref{bbkbkt}) and their stability for various
dimensionality goes beyond
the scope of this work. We shall analyze the flow in a reduced parameter space which is relevant for the 
issue of asymptotic safety in gravity. For $\mu_k, \kb_k\not=0$ we can consistently assume that 
$\nu_k=\kx_k=0$, as then the associated $\beta$-functions vanish. The reduced action (\ref{avac}) contains only the
Einstein and cosmological-constant terms. It must be emphasized that the theory we are considering is not dynamical gravity. 
The action (\ref{avac}) involves only one fluctuating scalar degree of freedom that has geometric origin. Despite the different 
nature of the theory, we find that the flows display striking similarity with what has been observed in the analysis of
gravitational theories.

The evolution of the couplings is described by eqs.~(\ref{bbmukt}), (\ref{bbkbkt}) with $\kxt_k=0$. We concentrate on the case $d=4$ 
which is closest to four-dimensional gravity. 
In order to make the analogy with gravity more apparent we define the {\it dimensionless} cosmological and Newton's 
constants through the relations
\be
\mut_k=\frac{\Lx_k}{8\pi G_k},~~~~~~~~\kbt_k=-\frac{1}{8\pi G_k}.
\label{dimens} \ee
Their evolution is given by
\begin{eqnarray}
\partial_t \Lx_k&=&-2\Lx_k+\frac{1}{6\pi}G_k(3-2\Lx_k)
\label{Lk} \\
\partial_t G_k&=&2G_k+\frac{1}{12\pi}\frac{G^2_k}{\Lx_k}(3-4\Lx_k).
\label{Gk}\end{eqnarray}
This system of equations has two fixed points at which the $\beta$-functions vanish: 
a) the Gaussian one, at $\Lx_k=G_k=0$, and b) a nontrivial one, at $\Lx_k=9/8$, $G_k=18\pi$.

\begin{figure}[t]
\begin{center}
\epsfig{file=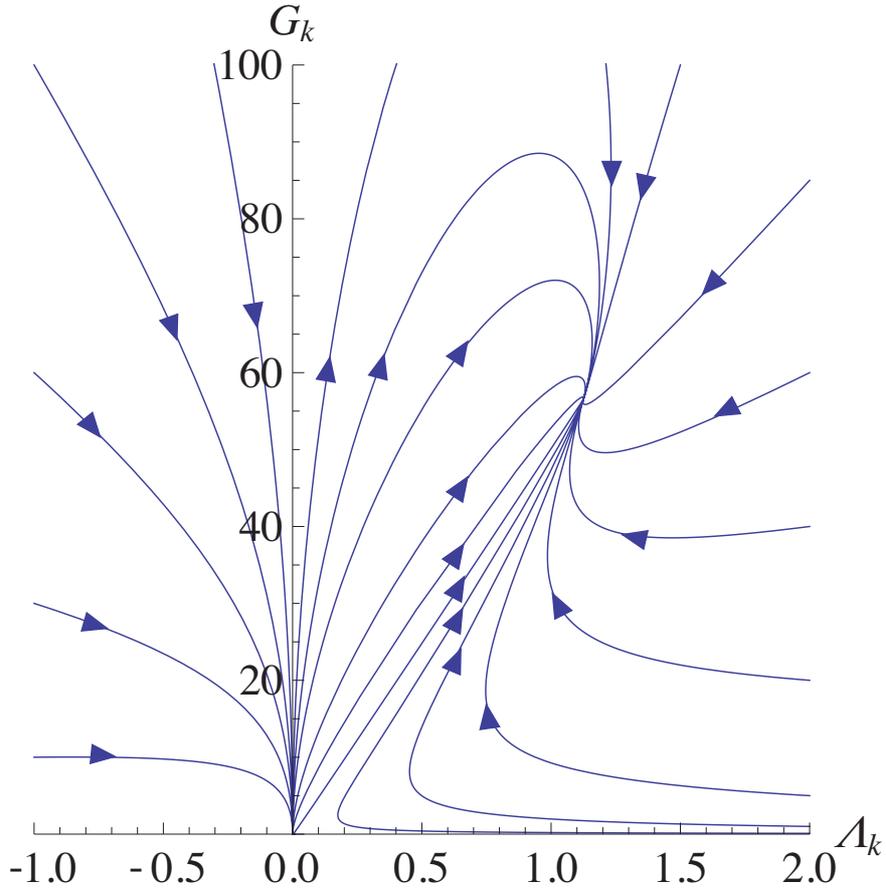,height=12cm}
\end{center}
\caption{The flows predicted by the evolution equations (\ref{Lk}), (\ref{Gk}).}
\label{flows}
\end{figure}

The evolution of the couplings is depicted in fig. \ref{flows} for increasing $k$. 
For $\Lambda_k>0$ the Gaussian fixed point is UV
unstable, while all flows converge to the nontrivial one, which is UV stable. 
This indicates that the nontrivial fixed point can provide a UV completion of the theory by 
allowing the limit $k\to \infty$ to be taken. The flows in the region $\Lx_k<0$ converge towards the Gaussian 
fixed point, which is now UV stable. The two regions, of positive or negative $\Lx_k$, are disconnected, as the $\beta$-function
of $G_k$ diverges on the line $\Lx_k=0$, while the flows are in opposite directions on either side of this line. 

The presence of the nontrivial fixed point and the form of the flows around it 
display a strong similarity with the corresponding flows for gravity in the Einstein-Hilbert truncation, in which only the 
cosmological and Newton's constants are retained \cite{asysaf}. In gravity the flows for $\Lx_k<0$ or for large positive $\Lx_k$
can display a strong sensitivity to the choice of the cutoff function.  However, their qualitative form in the vicinity of the fixed points
is stable and provides support for the asymptotic safety scenario, which assumes a UV completion of gravity through a nontrivial fixed point.
A nice feature of our flows is that they display stream lines connecting the region near the 
UV fixed point with the physical IR region in the limit $k\rightarrow0$.

An important conclusion of our study is that the asymptotic safety scenario can be realized even within 
scalar theories, which are in general much simpler to analyze. In this sense these results are similar to those obtained by considering the gravitational flows induced by matter fields in the large $N$ limit \cite{Percacci_2006}.  It must be emphasized, however, that the 
theory we are considering has an underlying gauge symmetry, the reparametrization invariance of the 
worldvolume, which must be preserved in the cutoff theory. In this sense it poses difficulties analogous to those encountered
when trying to preserve the general covariance of gravity.
 
\section{Conclusions}\label{sec6}

The focus of this work has been on understanding the effect of quantum corrections on the
structure of higher-derivative theories. Such theories are in general nonrenormalizable in the perturbative sense. 
For this reason we 
employed the ERG, which has the potential to reveal nonperturbative features, such as fixed points not easily 
accessible to perturbative methods. On the other hand, it must be kept in mind that 
the ERG approach relies heavily on the use of truncated
versions of the effective action, which may not capture all the physics.

The analysis of a general higher-derivative theory would involve too many
parameters. For this reason we limited our discussion to the class of theories that describe 
$d$-dimensional fluctuating branes within a bulk space of $d+1$ dimensions. The physical degree of freedom 
is the position modulus $\pi$ of the brane, which can be viewed as a scalar field of the worldvolume theory.
The structure of the Lagrangian density is constrained by the reparametrization invariance of the brane worldvolume.
The various terms correspond to geometric invariants, involving the extrinsic and intrinsic curvatures of the brane
expressed in terms of $\pi$. 

In the nonrelativistic limit the classical brane theory can reproduce the structure of the Galileon theory \cite{dbigal}. 
An important question is whether this feature remains valid at the quantum level as well. We found 
evidence that quantum corrections spoil the correspondence. They generate a geometric term in the brane theory
$\sim K^2$, where $K$ denotes the trace of the extrinsic curvature. 
Even if the term is absent at the
classical level, it will appear upon renormalization. In the nonrelativistic limit this term becomes 
$\sim \pi \Box^2 \pi$, a contribution not
present in the Galileon theory. On the other hand, the analysis of the quantum corrections to the 
Galileon theory through the use of dimensional regularization indicates that the lowest-order correction is
$\sim  \pi \Box^4 \pi$ \cite{quantum}. The discrepancy can be resolved by noting that the ERG analysis employs an 
explicit cutoff as a regulator of momentum integrals. For this reason it is sensitive to corrections with
quadratic or quartic divergences. The term $\sim \pi \Box^2 \pi$ is induced by a correction with a quartic
divergence, which is not visible through dimensional regularization. 
It must be noted that our conclusion does not depend on the specific choice of the infrared cutoff that we 
employ in the context of the ERG, and is, therefore, ERG-scheme independent.

We considered the action of eq.~(\ref{avac}), written in terms of geometric invariants. These can be
expressed through the position modulus $\pi$ according to eqs.~(\ref{sl})-(\ref{sk}).
The $\beta$-functions for the couplings of the theory are given by eqs.~(\ref{bmuk})-(\ref{bkbk}). They
form the main result of this work. 
For the particular choice (\ref{cutoff}) for the cutoff function, the $\beta$-functions can be written in 
the form (\ref{bbmukt})-(\ref{bbkbkt}), without an explicit reference to the running scale $k$.  
In an approximation consistent with these equations, we considered a truncation of the action that preserves only
the cosmological-constant and Einstein terms. Despite the similarity with 
dynamical gravity, the theory has only one fluctuating scalar degree of freedom.
It is remarkable, therefore, that the most prominent feature of the flow diagram is 
qualitatively similar to the one in the asymptotic-safety scenario for gravity. 
There is an attractive UV fixed point, which can be employed in order to obtain a UV completion of the theory. 

The fixed points and the related flows predicted by eqs.~(\ref{bmuk})-(\ref{bkbk}) for various values of $d$ will be
the focus of future research. The coupling $\nu_k$ can be consistently 
set to zero if we assume a symmetry in the fluctuations on either side
of the brane. The reduced system involves three couplings ($\mu_k$, $\kx_k$, $\kb_k$) and possesses novel fixed 
points. It forms a consistent framework in which to study the renormalization-group evolution of a higher-derivative
theory with nontrivial features. The analogy with the evolution of $d$-dimensional gravity is a very interesting
issue.

\section*{Acknowledgments}
We would like to thank S. Abel, J. Rizos and R. Percacci for useful discussions.
This research of N.T. has been supported in part by
the ITN network ``UNILHC'' (PITN-GA-2009-237920).
The research of N.T. has also been co-financed by the European Union (European Social Fund – ESF) and Greek national 
funds through the Operational Program ``Education and Lifelong Learning" of the National Strategic Reference 
Framework (NSRF) - Research Funding Program: ``THALIS. Investing in the society of knowledge through the 
European Social Fund".
The research of O.Z. is supported by the DFG within the Emmy-Noether program (Grant SA/1975 1-1).

\end{document}